\documentclass[final, 5pt, twocolumn]{elsarticle}

\usepackage{amssymb}
\usepackage{amsmath}
\usepackage{mathtools} 
\usepackage{etoolbox}
\usepackage{soul}
\usepackage{xcolor}
\usepackage{hyperref}
\usepackage{siunitx}
\usepackage{physics}
\usepackage{float}

\journal{}

\begin{document}
\makeatletter
\def\ps@pprintTitle{%
  \let\@oddhead\@empty
  \let\@evenhead\@empty
  \def\@oddfoot{\footnotesize\hfill\@date\hfill}
  \let\@evenfoot\@oddfoot
}
\makeatother

\begin{frontmatter}



\title{The effect of HVDC lines in power-grids via Kuramoto modelling}


\author[ek,bme]{Kristóf Benedek} 
\ead{benedek.kristof@ek.hun-ren.hu} 
\affiliation[ek]{organization={Institute of Technical Physics and Materials Science, HUN-REN Centre for Energy Research},
            addressline={Konkoly-Thege Miklós út 29-33, P.O. Box 49}, 
            city={Budapest},
            postcode={H-1525}, 
            state={},
            country={Hungary}}

\affiliation[bme]{organization={Department of Theoretical Physics, Institute of Physics, Budapest University of Technology and Economics},
            addressline={Műegyetem rkp. 3}, 
            city={Budapest},
            postcode={H-1111}, 
            state={},
            country={Hungary}}

\author[ek]{Géza Ódor} 
\ead{odor.geza@ek.hun-ren.hu}

\begin{abstract}
We present a numerical study on the synchronization and cascade failure behaviour by solving the adaptive second-order Kuramoto model on a large high voltage (HV) European power-grid.
This non-perturbative analysis takes into account non-linear effects, which occur even when phase differences are large, when the system is away from the steady state, and even during a blackout cascade. Our dynamical simulations show that improvements in the phase synchronziation stabilization as well as the in the cascade sizes can be related to the finite size scaling behaviour of the
second order Kuramoto on graphs with $d_s<4$ spectral dimensions. On the other hand drawbacks in the frequency spread and
Braess effects also occur by varying the total transmitted power at large and small global couplings, presumably when the
fluctuations are small, causing a freezing in the dynamics.
We compare simulations of the fully AC model with those of static or adaptive High Voltage Direct Current (HVDC) line 
replacements. The adaptive (local frequency difference-based) HVDC lines are more efficient in the steady state, at the
expense of very long relaxation times.  
\end{abstract}




\end{frontmatter}



\section{Introduction} \label{sec:intro}

Power-grids make up one of the most important infrastructures of technological
civilization as they distribute the energy from resources to consumers.
They are organized into the largest man-made synchronous machines as they
are based on oscillatory elements, related to the traditional rotating generator
sources and mechanical machines. 

However, nowadays this is about to change with the extension of renewable resources 
and other consumers which cannot be considered simple oscillators (see \cite{ENTSOEPLANNING2025})
To balance supply and demand, highly synchronized, continent-sized grids have 
been created, in which the phase differences drive the flow of energy from one node to the other. 
This permits describing classical power-grids by the so-called swing equations~\cite{Grainger1994}, 
equivalent to the second (or higher) order Kuramoto equations \cite{Filatrella2008}, 
exhibiting inertia. 
Inertia traditionally comes from rotating masses. By introducing elements based on new technology (e.g. renewable energy sources), direct current (DC) lines, etc.), this inertia is being reduced.
Therefore, power grids become more vulnerable, since with the reduction of the rotating inertia, smoothening perturbations in the power flow and frequency 
is decreasing. Because of this, instabilities and abrupt changes are not damped spontaneously.
The instabilities can arise from multiple sources, be it a change in demand and supply in different regions~\cite{pere_transmission2025}, operator mistakes, system failures, solar flares 
and many more. Due to the integration of renewables, the system is also subject to the big energy
fluctuations due to weather, amplified by climate change~\cite{Stürmer2024}. 
Large fluctuations in the frequency lead to the disintegration of the synchronization, 
cascade failures and blackouts~\cite{spainB25}.

There are several attempts to mitigate these fluctuations: addition of
inertia at critical places~\cite{Park_2025}, 
usage of adaptive inertia~\cite{PRXEnergy.3.033003}, and another solution is the application of direct current (DC)
lines partitioning the power-grid into smaller parts~\cite{Gomilla2025}. 
This technology is well-known and used in the case of high-power and long-distance (typically seabed) cables, 
as energy can be transmitted more efficiently over them than via alternating current (AC) ones.
Modification of power-grids do not always lead to improvements even if they are planned to be
like that. It is well known that Braess paradox \cite{Braess1968,Braess2005} can occur by transmission line improvements \cite{Schafer2022, CSFcikk25}. Here we investigate this question as the function of different methods as the function of maximal transmitted power.

Synchronization transitions are related to the spectrum of the Laplacian matrix of complex networks. The normalized Laplacian matrix defined as $L_{ij}=\delta_{ij}-A_{ij}/k_i$, where $\delta_{ij}$ denotes the unit, $A_{ij}$ the adjecency matrices and $k_i$ is the degree of node $i$. As the linearized Kuramoto equation describes a random walk, the eigenvalue spectra of $L_{ij}$ encodes the dimensionality information of a network related to synchronization as well~\cite{burioni1996}. In particular the density of which is characterized by the following scaling behavior \cite{burioni1996,millan2018}
\begin{equation}
    \rho(\lambda)\simeq \lambda^{d_s/2-1}
    \label{eqs:dsl}
\end{equation}
for $\lambda\ll 1$, where $d_s$ is the spectral dimension.

We know that power-grid topologies are characterized by rather low spectral dimension ($d_s$)~\cite{chimera_odor}. At $d_s < 4$ dimensions phase synchronization phase transition of the second-order Kuramoto model (KM) cannot happen and even the frequency entrainment phase transition does not occur for $d_s<2$. That does not exclude partial synchronization with a crossover; however, that means that synchronization coupling strength $K_c$ grows and diverges with the system size by the laws shown by Ref.~\cite{Odor2023}. As a consequence, smaller systems exhibit smaller $K_c$ and the DC separation of a power-grid graph must lead to sub-systems with higher synchronization values for a given $K$.

In this study, we show by solving the adaptive KM on a real European high voltage HV power-grid 
(from 2016), how the application of different link DC methods, on lines connecting the Scandinavian 
peninsula to the rest of Europe, modify the synchronization and cascade failure behavior. 
Our work and results provide a complement to available ones by electrical engineering (EE) from
a deeper, physicist based point of view.

\section{Previous works} \label{sec:prevw}

There seems to be at least two sides of the HVDC separation advantages:
\begin{itemize}
    \item reduction of cascade failures, which can be taken into account even by DC load models.
    \item improvement of synchronisation measures (and stability), which can be modelled by AC-DC models.
\end{itemize}

At first glance one may assume that large power systems allow the possibility to balance between energy demand and resources,
thus they provide more stability for power-grids. However, recent studies have demonstrated that this is not always the case,
and segmentation of power systems can be advantageous \cite{Does-size-matt,Why-southern}.
An early study, based on a risk assessment method, has shown that HVDC segmentation can significantly reduce the risk of
cascading failures~\cite{Mousavi2013}. This model is based on the simple fact that cascading outages and collapse will
be limited to only one segment, which reduces the risk of widespread blackout. Several variants of
simple DC/EE models have been designed and investigated to prove this. 


In the study of synchronisation and stability in large power networks, prior work mainly relies on two modelling paradigms: \textit{load-flow/OPF-based methods} and \textit{OPA-type cascading-failure models}. Optimal power-flow (OPF) formulations based on AC or DC load flow remain the standard for characterising feasible steady-state operating points under network constraints \cite{kundur1994,SauerPai1998}. They also provide the operating equilibria from which linearised small-signal stability or nonlinear swing-equation studies are conducted. Such OPF-derived operating points have been extensively used in research on grid synchronisation, stability margins, and vulnerability to large transfers in meshed transmission systems \cite{dorfler2013synchronization,motter2013spontaneous, pere_opa_model}.

Complementing these approaches, the \textit{ORNL–PSERC–Alaska (OPA)} model \cite{OPA-Carr-2004,OPA-improved-Mei} offers a coarse-grained representation of cascading failures based on a DC load-flow approximation. Its stochastic outage rules and slow-timescale adaptation enable the study of large blackout statistics and systemic risk. The recent application of OPA to the European transmission grid \cite{Gomilla2025} demonstrates how splitting the system into asynchronously coupled regions via HVDC corridors can reduce large-scale cascading risk in an economically favourable manner. However, because OPA relies on a *fully DC approximation*, it cannot capture angle-dynamics, reactive-power effects, or loss-of-synchrony phenomena in the AC system.

Across both OPF-based and cascading-risk models, \textit{HVDC interconnections} are typically represented as controllable power-transfer elements that fully decouple AC phase angles. In steady-state load-flow and OPF formulations, an HVDC link is usually modelled as two converter terminals enforcing specified injections, accompanied by converter losses and operational limits \cite{JovcicHVDC2019}. In stability-focused studies and synchronisation modelling, HVDC links appear as idealised controllable injections that do not transmit inertia or synchronising torque, thereby enabling regional decomposition of the AC grid. This abstraction captures the essential feature that HVDC allows power sharing without enforcing frequency or phase synchrony, making it a suitable tool for mitigating cascading failures or intentionally segmenting large synchronous areas~\cite{controlled_islanding,Energies2024HVDCdroop}.

Another recent work has analysed the synchronous AC grid of continental Europe under
tunable large inter-regional power flows \cite{pere_transmission2025}. This study
starts from the AC swing equations and applies a linear stability approximation to
identify critical lines in the EU and develops a numerical approach to force the splitting 
of the AC grid into disconnected areas. This is basically also a load‑flow‐driven model in which
"HVDC‑like" decoupling is emulated by cutting lines whose phase difference saturates, i.e.\ $|\theta_i - \theta_j| > \pi/2$.

Complementary to segmentation, modern control seeks to compensate lost inertia and stabilize frequency. Adaptive (time‑varying) synthetic inertia was shown to suppress coherency oscillations and improve stability margins in high‑renewables grids~\cite{PRXEnergyAdaptiveInertia}. There is also progress on where to place inertia: optimization on second‑order Kuramoto dynamics with threshold‑based tripping suggests that \emph{where} inertia is added matters as much as \emph{how much}, with peripheral placements often outperforming central ones~\cite{Park_2025}.

We will investigate this problem via the extension of the second order Kuramoto (AC–DC) 
model with an underlying line‑failure dynamics, which can treat both sides, 
without using a linear approximation. This can be interesting in light of the 2025 Iberian blackout,
which has renewed attention to voltage/frequency stability and the role of inter‑area separation and controls at scale~\cite{ENTSOE_Iberian_Blackout_2025}.

\section{Models and Methods} \label{sec:methmod}
\subsection{Dynamical simulations of the AC power-grid via Kuramoto model} \label{subsec:general}

The dynamics of the power-grid built up for mechanical elements 
(e.g.~rotors in generators and motors) can described by the swing equations~\cite{Grainger1994}.
It is equivalent to the Kuramoto equation of second type~\cite{Filatrella2008}
possessing inertia, which can be set up for a network, containing $N$ oscillators 
with phase variables $\theta_i(t)$. We use here a specific form, including 
dimensionless electrical parameters and approximations for the unknown 
quantities as in~\cite{Taher2019,Odor2018,Odor2022,Hartmann2024},
\begin{equation}\label{eq:kur2eq}
{\ddot{{\theta }}}_{i}+\alpha {\ }{\dot{{\theta}}}_{i}=P_i
+\frac{{P}_{i}^{max}}{{I}_{i}{\ }{\omega }_{G}}{\ }\sum
_{j=1}^{N}{{W}_{\mathit{ij}}{\ }\sin \left({\theta }_{j}-{\theta
}_{i}\right)} \ .
\end{equation}
Here $\alpha$ denotes the damping factor, which describes
the power dissipation, or instantaneous feedback~\cite{Odor2020},
$K:=P_i^{max}$ is a global control parameter, related to the
maximum transmitted power between nodes, $I_i=I$ is the inertia and
$\omega_G$ describes the generator frequency, which are assumed to be
constants in the lack of our knowledge. 
$W_{ij}$ denotes the weighted adjacency matrix of the network, containing 
admittances, calculated from line susceptances as described 
in~\cite{Hartmann2024}.
The fixed external drive, denoted by $P_i:=\omega_i^0$, corresponds 
to the the self-frequency of the $i$-th Kuramoto oscillator,
describes the input-output powers of the nodes.

Here, in the absence of our knowledge of the input-output nodal powers,
self-frequencies are set to be random variables, drawn from a zero-centered Gaussian
ensemble. Note, that the rescaling invariance of (\ref{eq:kur2eq}) allows
to transform these in a rotating frame to obtain other, variable nodal power values
(see \cite{Odor2022}).
We also assumed that the initial frequencies of the nodes $\omega_i(0)$ are the
same as their self-frequencies: $\omega_i(0)=\omega_i^0$.
In this study the following damping parameter setting was used
for the dissipation factor $\alpha=0.4$, which is 
common in real power-grid models, when they are parametrized with real 
physical dimensions~\cite{CSFcikk25}, but this can also be transformed
via the rescaling invariance of (\ref{eq:kur2eq}) to other values.

To integrate the differential equations we used the adaptive
Bulirsch-Stoer stepper~\cite{boostOdeInt} in the case of the GPU code, 
while running on CPUs we used the adaptive step-sized fourth-order 
Runge--Kutta method~\cite{WattsShampine2004}.
The nonlinearity of (\ref{eq:kur2eq}) introduces a chaotic 'noise' 
even without stochasticity, and a desynchronization transition occurs
by lowering $K$ from high value. The solutions depend on the actual
quenched $\omega_i^0$ self-frequency realization and to achieve
reasonably small fluctuations of the averages of the measured quantities, 
we needed strong computing resources, generally parallel codes 
running on GPU HPC machines. As (\ref{eq:kur2eq}) exhibits a first order
transition with metastable solutions even in the long time limit, to obtain 
stronger synchronization solutions, the initial state was set to be phase 
synchronized: $\theta_i(0) \simeq 0$.

We measured the Kuramoto phase order parameter:
\begin{equation}\label{eq:ordp}
z(t_k) = 1 / N \sum_j \exp\left[i \theta_j(t_k)\right] \ .
\end{equation}
Since $z(t_k)$ is a complex-valued number, the phase alignment at a given time $t_k$ is given by its modulus:
\begin{equation}
    r(t_k) = \lvert z(t_k) \rvert \ .
\end{equation}
We also computed the universal order parameter
\begin{equation}
    r_{uni}(t_k) = 1/(\sum_{i,j}^N W_{ij}) \sum_{i,j}^N W_{ij} \cos(\theta_i-\theta_j) \ ,
\end{equation}
which describes the energy of classical rotator models
and the variance of the frequencies :
\begin{equation}\label{eq:FOP}
        \omega(t_k) = \frac{1}{N}  \sum_{j=1}^N (\overline\omega(t_k)-\omega_jt_k))^2  \ ,
\end{equation}
where $\overline\omega(t_k)$ denotes the spatial average at time $t_k$.
Sample averages for the Kuramoto order parameter:
\begin{equation}\label{KOP}
R(t_k) = \langle r(t_k)\rangle \ ,
\end{equation}
for the universal one:
\begin{equation}\label{eq:runi}
    R_{uni}(t)=\langle r_{uni}(t_k) \rangle \ ,
\end{equation}
and for the frequency variances
\begin{equation}\label{eq:KOP}
\Omega(t_k) = \langle \omega(t_k)\rangle \ ,
\end{equation}
were performed over hundreds of independent self-frequency realizations.
As the self-frequencies generate a quenched disorder variances of the order parameters
were also calculated without having stochastic noise. 
To save data storage and computing time the solutions were sampled at discrete, 
incremental time steps: 
$t_k = 1 + 1.08^{k}$, $k=1,2,3...$.
in accordance with the relaxation behavior worst near the synchronization
point $K_c$.

We have extended the numerical solution equation of motions with a threshold dynamics, 
to model overflow of power on the edges that were removed during the simulation, 
generating a cascade failure. 
This method is similar to what was published in~\cite{Schafer2018,Odor2022}.
Following a thermalization to steady states, chacarterized by $R$, $R_{uni}$, $\Omega$, 
initialized from phase-ordered states, which we inspected by plotting the
order actual parameters, the systems were perturbed by removing a randomly 
selected edge, in order to simulate a power line failure event. 
Following that, if the ensuing power flow on a line between neighbouring nodes 
was greater than a threshold:
\begin{equation}\label{eq:Fij}
        F_{ij} = | \sin(\theta_j-\theta_i) | > T\,,
\end{equation}
the line was regarded as overloaded and we removed this link from the adjeceny
graph permanently (in run). In connection with this we calculated the total number 
of line failures $\langle N_f\rangle$ of the simulated blackout cascades of each realization, 
corresponding to different $\omega_i(0)$. Following the cascade simulations 
we applied histogramming to determine the probability distribution functions (PDFs) 
of $N_f$-s.

We have investigated various modifications of European power-grids introduced in a previous work~\cite{Hartmann2024}. 
These graphs are based on data from 2016 for Europe (EU2016)~\href{https://zenodo.org/records/47317}{GridKit project}, which relies on the ENTSO-E's statistics for power generation and consumption, voltage levels; and data obtained from OpenStreetMap (.osm) files. These contain information on the topology, geographical coordinates of nodes, and lengths, types, voltage levels of cables. The EU2016 network contains $N=$ \num{13478} nodes and $E=$ \num{18393} bi-directional edges and a hierarchical modular structure~\cite{Odor2022}. Since we use publicly available .osm files, this implies, the data may be badly labeled, meaning there is a possibility of having missing data for certain grid elements, resulting in an incomplete dataset in terms of parameters. To resolve the problem, we made assumptions in~\cite{Hartmann2024} to substitute the missing data in order to obtain a fully weighted network. 
The spectral dimension of the weighted graph was found to be $d \simeq 2.34$ ~\cite{chimera_odor}. 
In~\cite{Hartmann2024} a complete graph invariant and community structure analysis was also presented.

\subsection{Including HVDC lines} \label{subsec:hvdc}

In modern power systems, the use of HVDC transmission has emerged as a key solution for efficiently linking remote generation sites, interconnecting asynchronous grids and delivering bulk power via submarine or underground cables. Unlike the more conventional AC systems, an HVDC link carries electricity as direct current typically in the range of $\SI{100}{\kilo\volt}-\SI{800}{\kilo\volt}$ which offers a number of compelling advantages, such as lower transmission losses over long distances and the requirement of fewer conductors, since there are no three phases and no skin effect to worry about.

At the same time, HVDC cable systems are not without drawbacks. The terminal converter stations (AC$\leftrightarrow$DC and vice versa) are complex and expensive. For short transmission distances, the losses and cost of the conversion equipment may outweigh the savings in cable or line cost, meaning there is a so-called “break-even” distance beyond which HVDC becomes economical. Moreover, expanding DC systems into multi-terminal or meshed arrangements is still more challenging than with AC, because DC protection, switching, and converter coordination are more complex.

HVDC lines transmit power only; there is no phase or frequency information carried over. In the synchronization sense, they segment the network into distinct clusters. The key question arising when it comes to their modelling is how to implement regularization of the power. This can be done in sevaral ways~\cite{entsoe_hvdc_technical_paper, op_control_hvdc}:
\begin{enumerate}
    \item Converter control strategies
    \begin{enumerate}
        \item Voltage droop control: a proportional relationship between DC voltage and the required power.
        \item Constant power control: when constant power is needed, independent of the voltage.
        \item Deadband droop control: a mixture of the above two.
    \end{enumerate}
    \item Grid control strategies
    \begin{enumerate}
        \item Centralized voltage control: It is based on how the voltage source converter (VSC) HVDC point-to-point connection\footnote{Typically used to connect the DC grid to weak and passive systems making it the best choice when connecting to eg. offshore wind plants.} is operated~\cite{dc_grids_beerten_phd}.
        \item Distributed voltage control: Instead of having only one converter that controls the DC voltage in the grid as in centralized voltage control, distributed voltage control applies voltage droop control to several converters.
    \end{enumerate}
\end{enumerate}
Very often, droop control or constant power control is implemented. Droop control introduces adaptivity into the system. Since all these techniques depend on converters, which can be programmed,  power control could be influenced by voltage angle phase differences or frequency differences. We implemented several methods for this adaptivity, based on the frequency. We were inspired by the fact, that synchronous control in power grids is realized via frequency tuning~\cite{yan_freq_control}.

\begin{figure}[h]
    \centering
    \includegraphics[width=\linewidth]{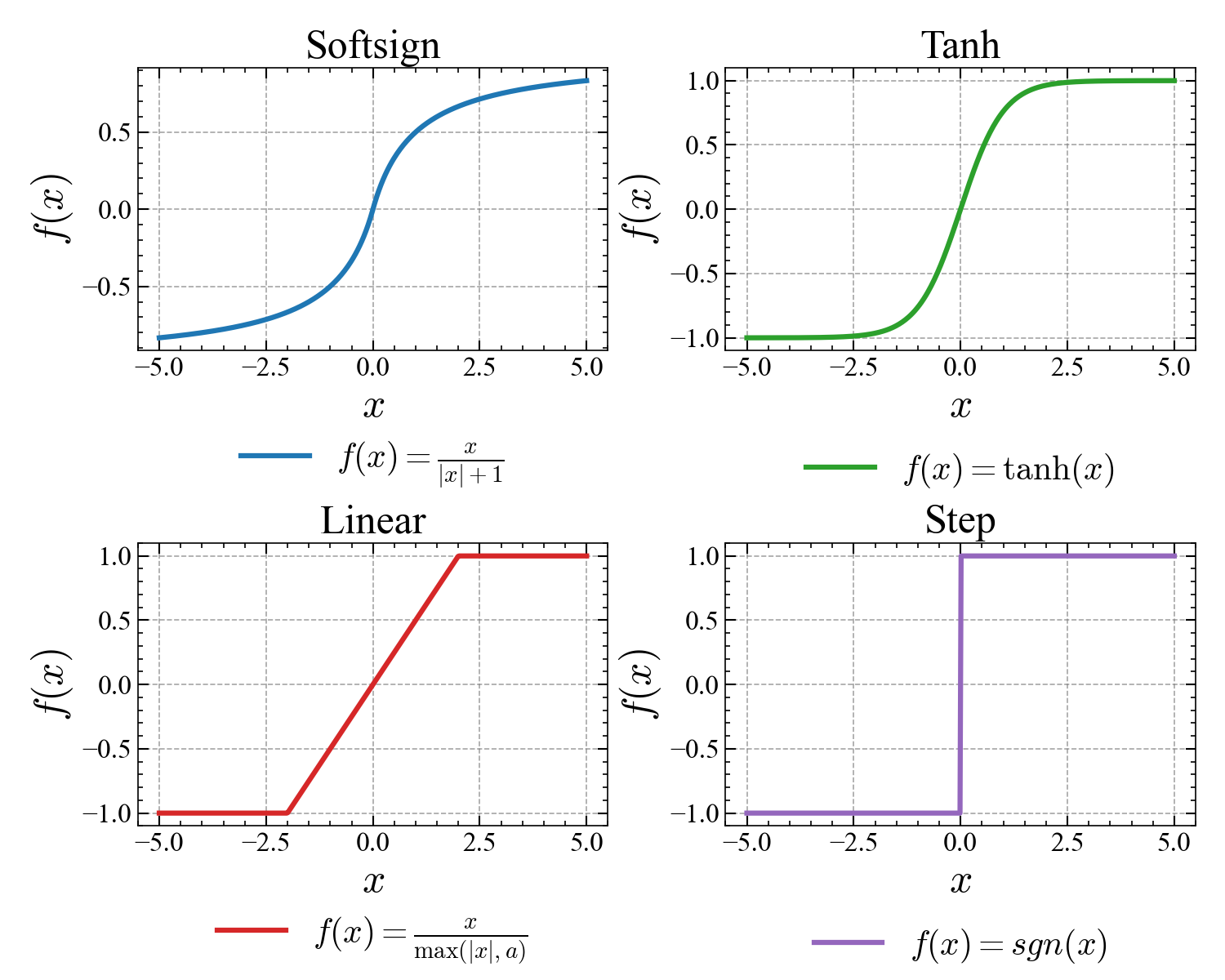}
    \caption{Different types of activation functions used in the adaptive HVDC modelling scenarios.}
    \label{fig:activations}
\end{figure}

We included the adaptive control in the Kuramoto model by modifying \eqref{eq:kur2eq} as follows:
\begin{align}
    \ddot{\theta}_i+\alpha\dot{\theta}_i&=P_i
+\frac{{P}_{i}^{max}}{{I}_{i}{\ }{\omega }_{G}}\sum
_{\substack{j, \text{ if edge is AC}}} W_{\mathit{ij}}\sin \left({\theta }_{j}-{\theta
}_{i}\right)\notag\\
&+\sum_{\substack{j, \text{ if edge is DC}}}f(\dot{\theta}_j-\dot{\theta}_i)\cdot 
D_{ij} \ .\label{eq:kureq2hvdc}
\end{align}
Here we introduced a new term, in which $D_{ij}$ is the strength of the HVDC link, proportional to the maximal transmissible power through it and $f(x)$ is a so-called \textit{activation function} (c.f. Fig. \ref{fig:activations}), normalizing 
the frequency difference between $-1$ and $1$. This adds a possibility for setting the power flow across DC lines and can be further combined with static power control. We expect that via this term the grid gains a better ability to self-organize toward synchrony, visible in achieving higher $R$ order parameter values, preferably lower $\Omega$ frequency spread and cascade sizes $\langle N_f \rangle$.

On the other hand, constant power control was also implemented and tested. To model this approach to HVDC lines, we removed these links from the graph by simply setting their weight $D_{ij}=0$, realising the grid segmentation as discussed before, and then we assumed that at the endpoints of these edges the nodes receive a specific input/output ($Pd_i = -Pd_o$) power(static HVDC).

\begin{figure}[H]
    \centering
    \includegraphics[width=0.75\linewidth]{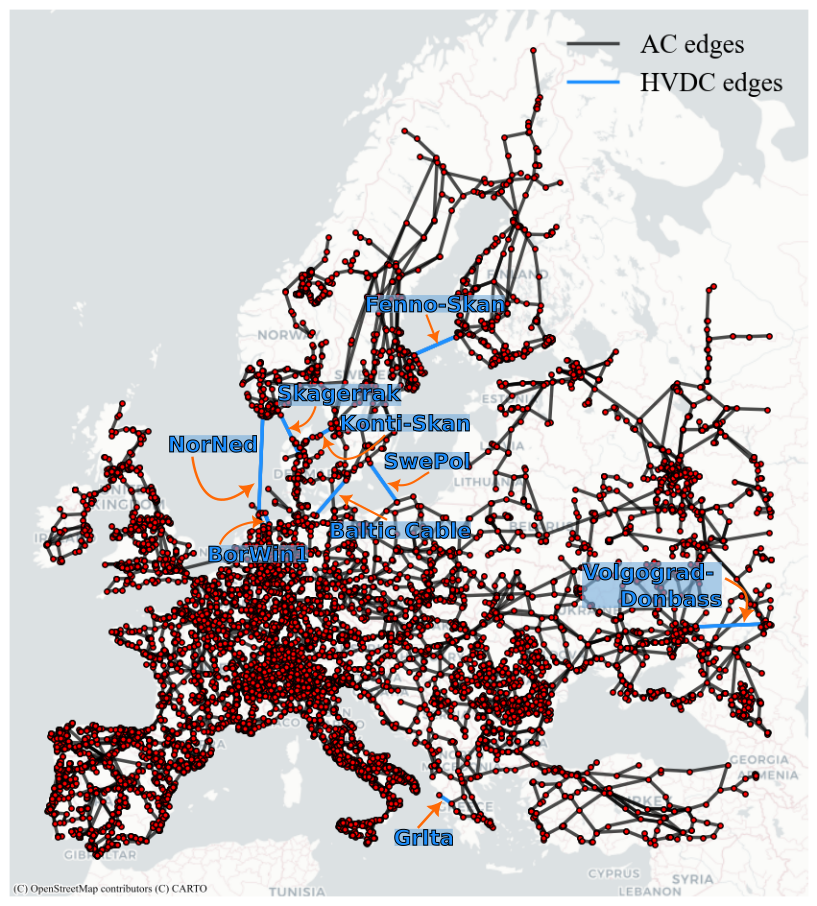}
    \caption{HVDC cables in the continental European grid based on the 2016 data in GridKit. While the Baltic area is quite well documented, the Southern-Europe part is less so. In this work only the Baltic seabed cables were turned from AC to DC. (c.f. \href{https://www.hvdcworld.com/hvdc-map?authReq=true&lat=45.266983&lng=6.752306&zoom=8.00&id=250}{HVDC World map}).}
    \label{fig:hvdc_cables_e16}
\end{figure}

In order to investigate the HVDC links across Europe, the first task was to identify the specific lines 
that were present in the ENTSO-E 2016 database, with their transmitted power. 
We decided to see the effects of replacing the seabed lines below the Baltic Sea. 
This amounts to $36$ HVDC cables of the network, which is just $0.2\%$ of the total,
still has a large effect of decoupling the Scandinavian peninsula from the rest of the continent
as shown on Fig.\ref{fig:hvdc_cables_e16}.

It is crucial to acknowledge that in the absence of real power values of the EU2016 grid 
the $Pd_i$ had to be properly normed to be in accordance with the randomly chosen generator
and consumer values $P_i=\omega_o^0$, a zero centered, random Gaussian values with unit variance
($\sigma=1$). For the normalizing purpose of the HVDC node values, we used the nodal power
values from \cite{Hartmann2024} and assumed a correspondance between the $800 MV$ power 
with the $P_i = 3\sigma$. The real HVDC cables power values were obtained from the ENTSO-E 
database and the $Pd_i = -Pd_o$ were calculated and used in the Kuramoto equations
according to the above correspondance.

\section{Results} \label{sec:results}
This section presents our results for the EU2016 grid, where we implemented and tested various approaches to simulate HVDC connections. Earlier we used an edge weighting method described in \cite{Hartmann2024}. This involves grouping the edges according to voltage levels and assigning weights to the AC cables based on their voltages and susceptances. For DC cables, we assigned weights proportional to the total capacity.
\begin{figure}[h]
    \centering
    \includegraphics[width=\linewidth]{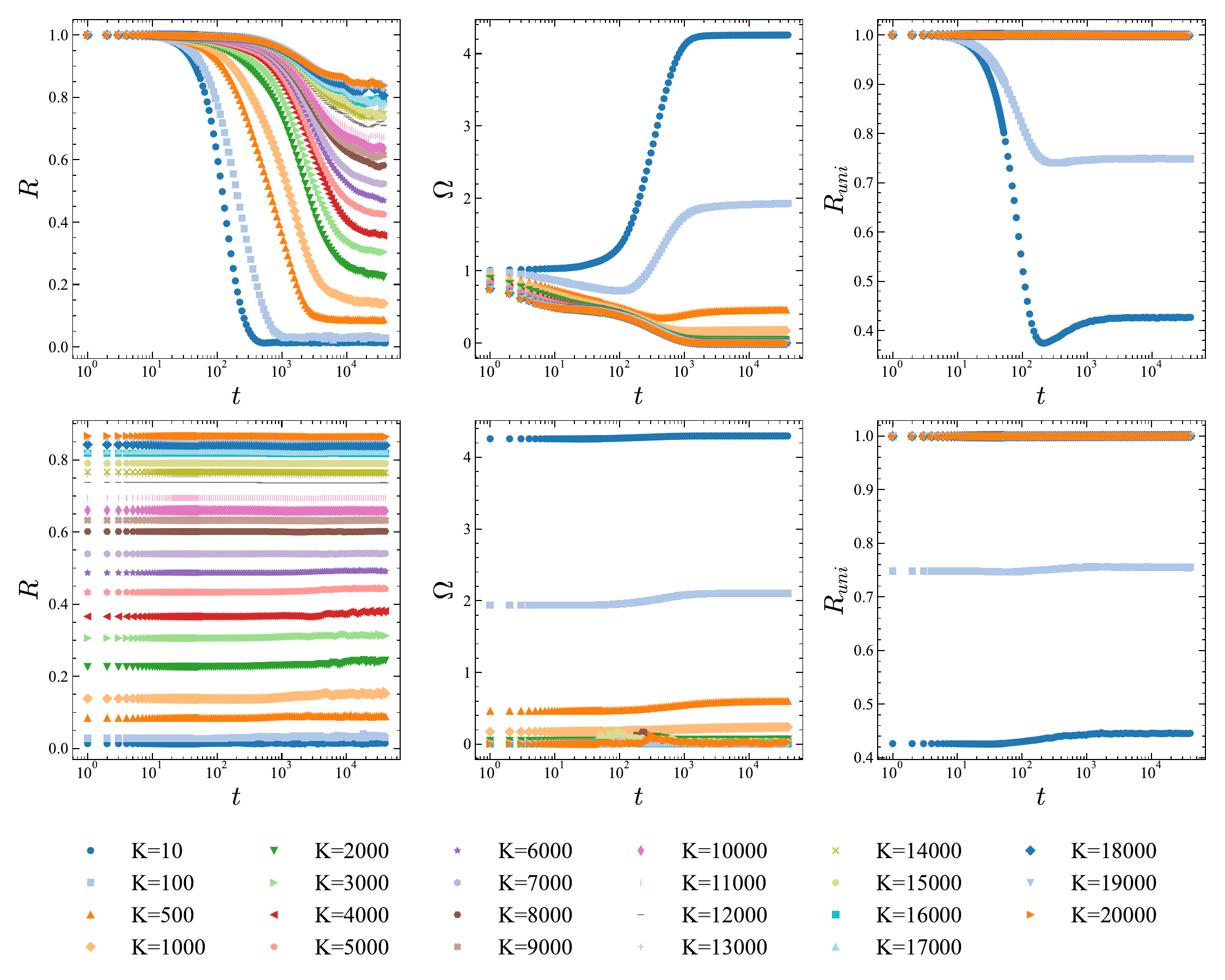}
    \caption{Example timeseries for HVDC modelling with $softsign(x)$ function as activation. The top row displays the thermalization phase, and the bottom one corresponds to the cascade part of the simulations. We are showing results for our main synchronization metrics, $R(t_k)$, $\Omega(t_k)$ and $R_{uni}(t_k)$, for various global coupling values, marked by different colors and symbols. Saturation to the steady state values requires long times, typically $10^4$ iterations and happens non-monotonically as we start from fully phase ordered initial states. The $R_{uni}(t_k)$ curves overlap above coupling $K=500$, being a hallmark of good local synchronisation .
    }
    \label{fig:timeseries_th_eu16}
\end{figure}

Investigating the EU2016 grid, we wanted to explore the impact of different activation functions (c.f. Fig \ref{fig:activations}) and their parametrisations on the synchronisation. In the domain of ENTSO-E, the power flow is regulated by Frequency Containment Reserves (FCR). FCR reacts to short-term frequency imbalances in the grid, the complete activation of the frequency reserve must be available within \num{30} seconds and cover a period of 15 minutes per incident according to ENTSO-E standards. Since HVDC elements can be viewed as building blocks segmenting the grid, they should be of key importance to balance out the power. However, the way they react to the frequency imbalances is encoded in the activation functions. In the electrical engineering context, FCR is realised via a linear function (Fig. \ref{fig:activations}, bottom left panel), that activates in a certain frequency band. The drawback is that this band has to be predefined and is arbitrary when it comes to theoretical investigation and modeling. Using $\tanh(x)$, $softsign(x)$ or $sign(x)$ functions has the natural benefit of not needing any additional parametrisation. It has to be mentioned, that the $sign(x)$ function can cause very large transients towards the steady state, (see Fig.\ref{fig:timeseries_th_eu16}) since at balance, the frequency difference sign can tip from negative to positive at every iteration and vice versa, which can also lead to slow convergence of the adaptive solver. Since in the literature~\cite{yan_freq_control, hvdc_transmission_overview, op_control_hvdc} HVDC elements are viewed as fine-tuning tools toward achieving better power flow and synchronisation in the system, we test this theory via dynamical Kuramoto simulations.
\begin{figure}[h]
    \centering
    \includegraphics[width=\linewidth]{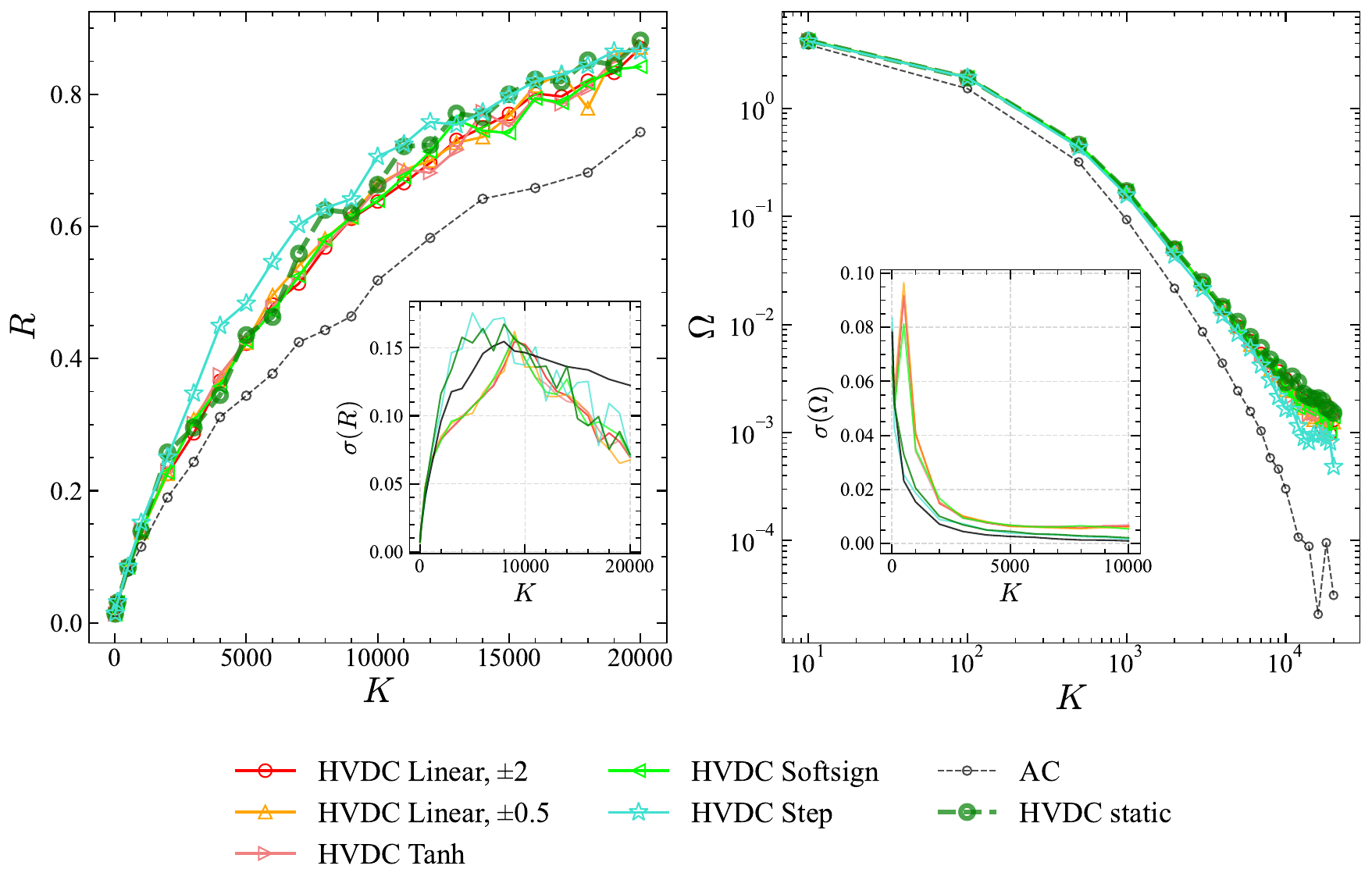}
    \caption{Comparison of the steady state results after the thermalization of the 
    fully AC and different HVDC approaches as the function of global coupling
    $K$ in case of the EU2016 network. The inset shows the standard deviation of these quantities.
    Notice the fact that even static power input/output at DC buses helps the synchronisation to 
    improve. On the other hand, the frequency spread is increasing, which can be understood, since
    DC connections do not transfer phase information.}
    \label{fig:therm_metrics_eu16}
\end{figure}
In Fig. \ref{fig:therm_metrics_eu16}, we show our results in the steady state after integrating Eq.\eqref{eq:kureq2hvdc} for \num{10000} 
iteration steps in cases of full AC, HVDC static, linear, $\tanh(x)$ and $softsign(x)$ activations, and for $t_{max}=300.000$ for the $sign(x)$ case. 
In the latter case, a much larger time was needed because of the aforementioned long transients. Nonetheless, each of the methods helps the synchronisation to improve across the grid. In the case of the full AC model, the half $R(t)$ synchronisation value is achieved at around coupling $K=7000$. 
Simulations for the linear, $softsign(x)$ and the $\tanh(x)$ HVDC activations result in a peaks in the $\sigma(R)$ at $K_c=9000$, hallmark of the phase synchronization transitions at larger global power flows. On the other hand, the $sign(x)$ and the static DC
functions decrease the peak locations to $K_c \simeq 5000$. Sample averaging was done over \num{200} random, independent self-frequency realizations.

On the second subplot of Fig. \ref{fig:therm_metrics_eu16}, we show the frequency spread results. As expected, in all of the HVDC modelling cases, this displays a higher value as compared to the full AC model. This can be understood, since DC connections do not transfer any phase and frequency related information between the two parts of the grid, only power. 
The total frequency distribution of the separated sub-networks can exhibit multi-peaks~\cite{Hartmann2025}. 
It was also shown in \cite{Odor2023} that the frequency entrainment point $K'_c(N)$ also grows with the system size, and since $\Omega(N,K)$ is a monotonous function, for the sub-networks the $\Omega(N_{sub},K) > \Omega(N,K)$ relation holds.
Note, that we don't see peaks of the $\sigma(\Omega)$ curves because frequencies exhibit first order synchronization transition.
We did not see much effect of HVDC lines with respect to the full AC on the $R_{uni}$ steady state results, thus we omitted to plot this.

\begin{figure}[H]
    \centering
    \includegraphics[width=\linewidth]{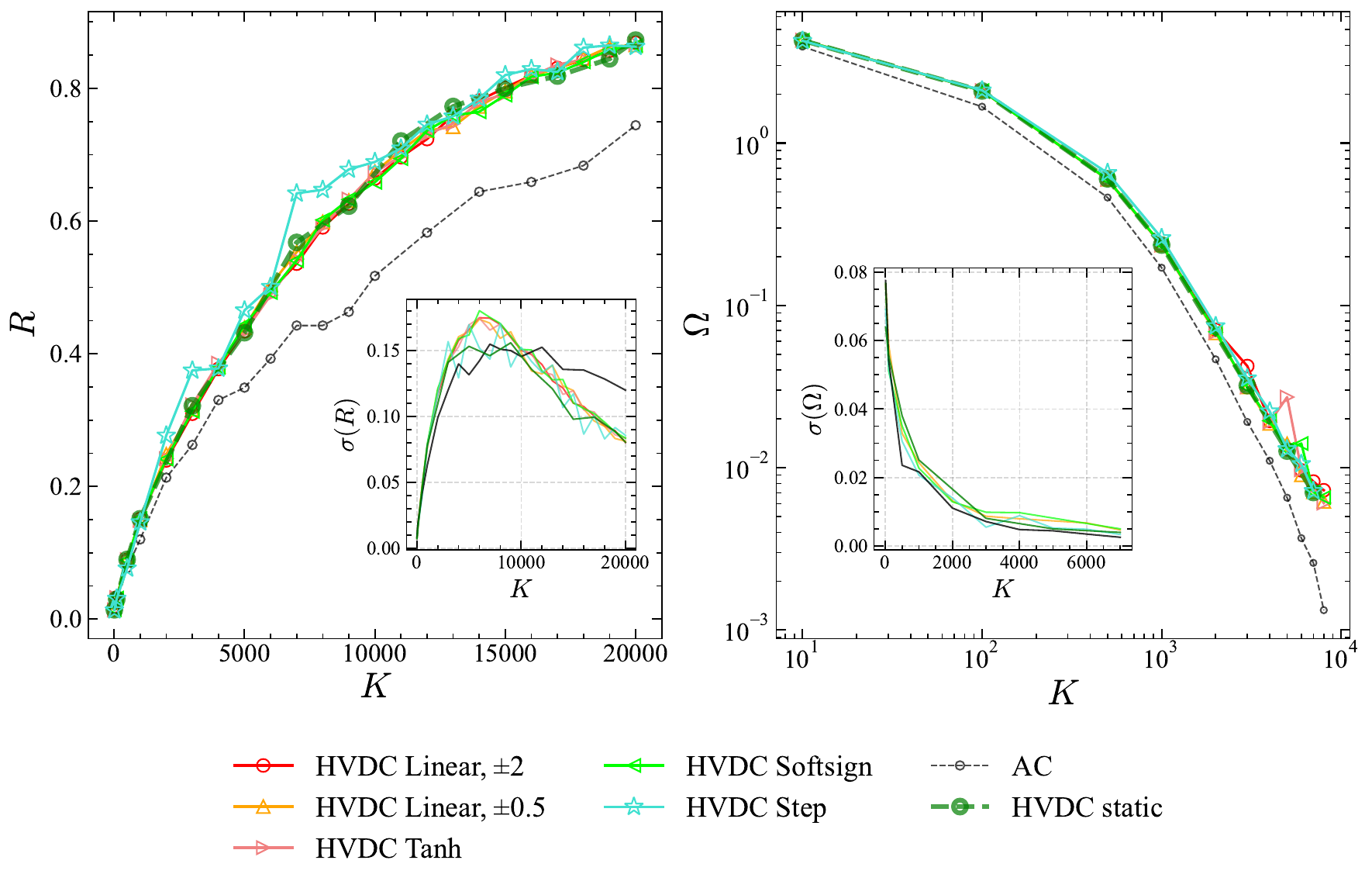}
    \caption{Comparison of the dynamical cascade simulation results of the 
    fully AC and different HVDC approaches as the function of global coupling
    $K$ in case of the EU2016 network. The inset shows the standard deviation of these quantities.
    Even during cascade simulation, a higher synchronisation is achieved when we account for the HVDC connections; however, the different approaches separate less than in the steady state (cf. Fig. \ref{fig:therm_metrics_eu16}). In terms of frequency spread the full AC model yields better results. This can be understood by the aforementioned fact that DC connections do not carry phase or frequency information.}
    \label{fig:casc_metrics_eu16}
\end{figure}

In the second part of our simulations, we created cascade events in the grid by manually cutting one of the randomly selected link in the steady state. This can lead to a domino effect of overloading lines by the condition (\ref{eq:Fij}, with $T=0.99$ Note, these power-grid models exhibit rather strong 
inherent instabilities in the steady states for moderate couplings, thus we applied this high threshold value.
Again, in the literature, it is claimed that HVDC links reduce the cascade risk, since they can respond rapidly to frequency oscillations and prevent failure propagation.

Like in case of the thermalization the dynamical cascade simulations were run until a steady 
state has been reached, determined by the order parameter differences, as shown on Figs. \ref{fig:timeseries_th_eu16}, \ref{fig:casc_metrics_eu16}.
While the static HVDC lines did not influence the relaxation times significantly, the adaptive ones can increase them drastically, making them \num{3} orders of magnitude larger. This is a natural consequence of the modelling. Taking for example, the linear activation function, as we decrease the interval for the linear part, we are turning the continuous equations into discontinuous ones, something that hinders all integrators in numerical methods.  
\begin{figure}[H]
    \centering
    \includegraphics[width=\linewidth]{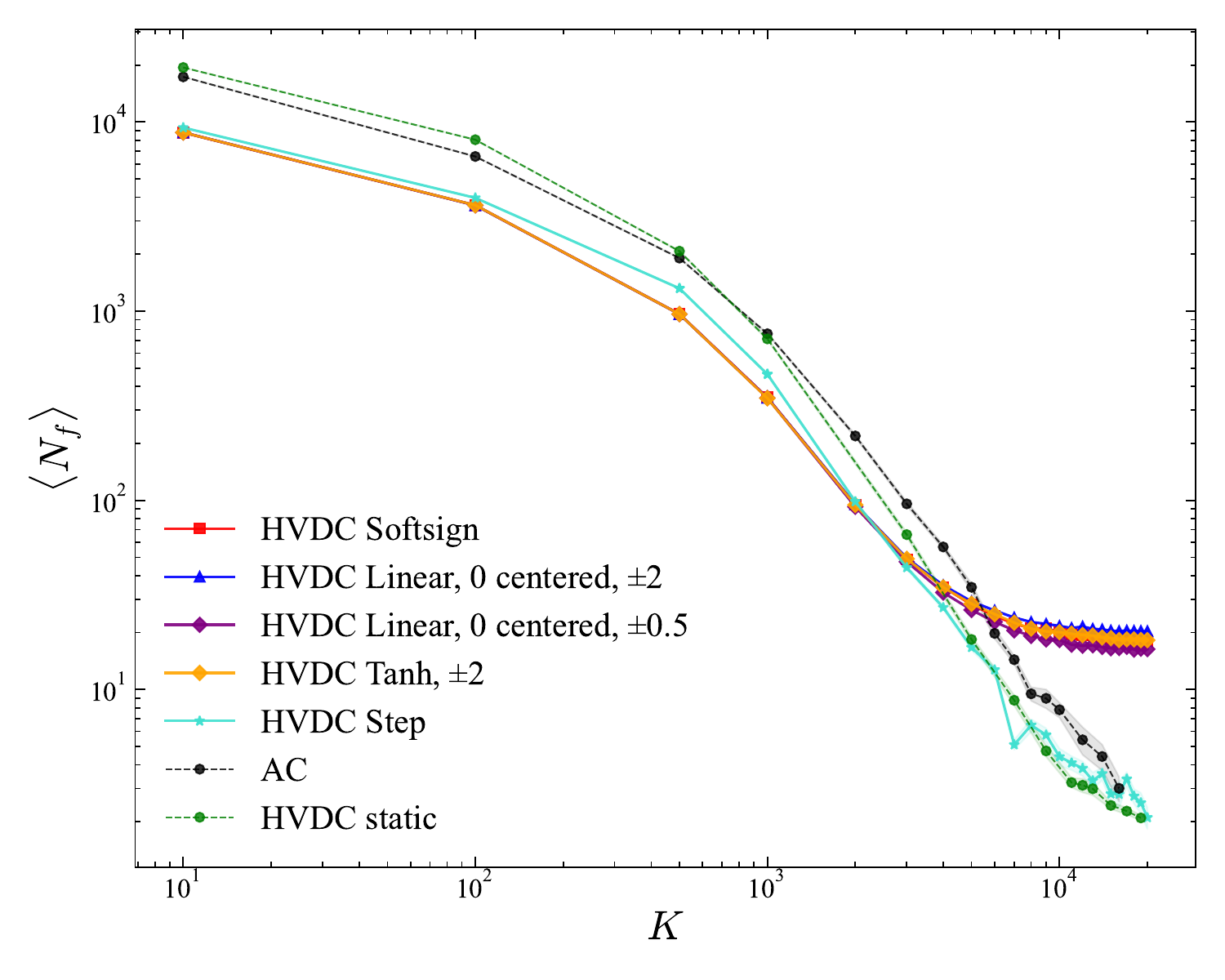}
    \caption{Cascade size distributions for EU2016 with different types and valued activation functions in the HVDC cable terms compared to previous results, full AC simulation and static HVDC modelling. The shaded region represents the standard error corresponding to each case. Choosing the static modelling or $sign(x)$ as activation function, results in a stabilizing, cascade-reducing effect for all investigated $K$ values. Other methods display a crossover behaviour, namely, they reduce the size distributions for lower couplings, but can't mitigate the failures for higher values. 
    }
    \label{fig:nf_eu16}
\end{figure}


We investigated the average size distribution of the cascade $\langle N_f \rangle$, as the function of the global coupling, for the different modelling scenarios, displayed on Fig. \ref{fig:nf_eu16}. We can see that linear activation, $\tanh{x}$ and $softsign(x)$ stabilizes the grid for $K$ values up to \num{6000}, but after that, they run into a saturation and can not mitigate the cascade anymore. 

In the case of the linear activation function (c.f. Fig. \ref{fig:activations} bottom left panel), there is a possibility to vary the interval of the linear activation part. We kept the functions $0$ centered, since the mean frequency, \SI{50}{\Hz}, can be transformed out from the Kuramoto equations \eqref{eq:kur2eq} and \eqref{eq:kureq2hvdc}, but we experimented with various interval sizes. We display results only for some of the interval sizes in Fig. \ref{fig:nf_eu16}, $\pm2,\;\pm0.5$, but the dynamics were checked for other values, e.g. $\pm1,\;\pm0.25,\;\pm0.1$, too. The general conclusion of these results is that by decreasing the size of the interval (approaching the edge-case step function), we can push the saturation plateaus of $R(t)$-s and $N_f(t)$ lower, decreasing the cascade sizes in the system.

In the case of $sign(x)$ and static DC models, we see a stabilizing effect throughout the whole investigated coupling regime.
%
%

\section{Conclusions} \label{sec:conc}

In conclusion, we presented dynamical cascade failure simulations on the EU2016 HV power-grid by
considering the effects of different HVDC separator lines between the Scandinavian Peninsula 
and the rest of the continent. Our method enables us to follow the evolution of synchronization
behaviour deep in the non-linear regime permitting the determination of the steady state 
following a blackout cascade.

The order parameter and the cascade size results could be interpreted by the knowledge of the 
second order Kuramoto model on graphs with spectral dimension $2< d_s<4$, where the synchronization crossover point moves to higher couplings as the function of size. 
As we replaced certain ($0.02\%$) of the AC connections, by DC ones using sources and 
sinks at their end nodes we separated the graph into two smaller sub-graphs 
Both the Kuramoto and the frequency order parameters move to higher values for a given $K$, 
which provides better phase, but worse frequency synchronizations at a given global 
power-level in case of static HVDC. Furthermore, the cascade sizes decrease for intermediate $K$-s, where the synchronization transition generates stronger fluctuations. Similar advances were found in case of line additions to the EU2016 grid, close to the synchronization transition coupling~\cite{Odor2024}, where we argued about the positive effects of possible self-organized criticality. In the future, the effects of different additive stochastic noises should be investigated.

A comparison of static and different adaptive HVDC methods shows that the most sensitive, step-function, method provides the best results in the steady state, avoiding the frequency Braess paradox of static HVDCs for large $K$-s, but the relaxation time increases formidably as this method adds rapidly fluctuating contributions to the equations of motions breaking their continuous nature. Possible extension of this work would be the analysis of frequency distributions and comparison with real data.
\section*{Software and third party data repository citations} \label{sec:cite}

The network data used for this study was based on the following sources: SciGRID project, Prebuilt Electricity Network for PyPSA-Eur based on OpenStreetMap Data~\cite{prebuiltosm}, both being part of PyPsa-EUR~\cite{pypsa}. For part of the simulations on GPU we used the KuramotoGPU code developed by Jeffrey Kelling, and for on CPUs we have used the code written by Kristóf Benedek and available via GitHub~\cite{Benedek_kuramoto2_Simulation_code_2025}.

\section*{Acknowledgments}
We thank Jeffrey Kelling for creating and maintaining the KuramotoGPU HPC GPU code, \'Aron Weidinger and M. T. Cirunay for locating and calculating the EU2016 HVDC power lines, Pere Colet, Damia Gomilla and B\'alint Hartmann for the useful comments and discussions.
Kristóf Benedek acknowledges the support by the Doctoral Excellence Fellowship Programme (DCEP), founded by the National Research Development and Innovation Fund of the Ministry of Culture and Innovation and the Budapest University of Technology and Economics. Support from the Hungarian National Research, Development and Innovation Office NKFIH (K146736) is also acknowledged.

\section*{Contribution}

Kristóf Benedek was responsible of code development and implementation, running the simulations, visualization of results, and developing adaptive HVDC modelling theory. Géza Ódor was the main lead in result interpretation and theory development.


\appendix




\bibliographystyle{unsrt}
\bibliography{hvdc}{}

\end{document}